\documentstyle[epsfig, twocolumn]{esa_sp_latex}

\begin{document}
%

\parindent 0pt
\parskip 10pt plus 1pt minus 1pt
\hoffset=-1.5truecm
\topmargin=-1.0cm
\textwidth 17.1truecm \columnsep 1truecm \columnseprule 0pt 

\title{\bf LUMINOSITY TO THE EDDINGTON LUMINOSITY RATIO IN AGN}

\author{{\bf B.~Czerny$^1$, H.J.~Witt$^2$, P.T. \. Zycki$^1$} \vspace{2mm} \\
$^1$ Copernicus Astronomical Center, Bartycka 18,00-716 Warsaw, Poland \\
$^2$ Astrophysikalisches Institut Potsdam, An der Sternwarte 16, 14482 Potsdam, 
Germany}

\def\LE{L/L_{Edd}}

\maketitle

\begin{abstract}

  We discuss the optical/UV/Xray spectra of AGN within the frame of the
  corona model of Witt, Czerny \& \. Zycki (1996). In this model both the
  disk and the corona accretes and release energy through the
  viscosity. The relative strength the the disk and the corona is therefore
  determined by the model. Translated into the spectra, this model allows
  to predict the spectral index $\alpha_{ox}$, measuring the relative strenth
  of the big blue bump with respect to the hard X-ray power law. Comparison
  of the predicted and observed distributions of $\alpha_{ox}$ allow us to
  conclude that the $\LE$ ratio in quasars cover typically the range $\sim
  0.01 - 0.1$ and it is broader in Seyferts, $0.001 -  0.3$, or even higher.
  We identify Narrow Line Seyfert galaxies with high $\LE$ ratio object 
  although, in principle, the second branch of high big blue bump objects
  for  $\LE$ below 0.001 is also predicted. \vspace {5pt} \\


\end{abstract}

\section{INTRODUCTION}

Overall similarity between the spectra and variability of active 
galactic nuclei
(AGN) and galactic black hole candidates (BHC) indicate that the basis 
mechanism is the same in both kinds of objects and scales easily with the mass
of the central black hole.  If we
restrict our study to radio quiet AGN with broad emission lines we can neglect
the dependence of the observed emission on the inclination
angle of an observer with respect to the symmetry axis of the nucleus 
since we are dealing 
with objects oriented 'face on', according to the generally accepted 
unification scheme (see e.g. \cite{ant93}). 
It is therefore tempting to attribute the observed 
diversity among the sources (well expressed as the ratio of the big blue bump
to the uderlying power law, e.g. \cite{wf93})  to the range of values of the luminosity to the
Eddington luminosity ratio, $\LE$, 
with no additional parameters involved.

In order to determine $\LE$ one can
either estimate the bolometric luminosity
of an object and the mass of the central black hole separately or  adopt
a particular spectrum model based on these parameters and fit it to the data.

\begin{figure}
  \begin{center}
    \leavevmode
\epsfig{file=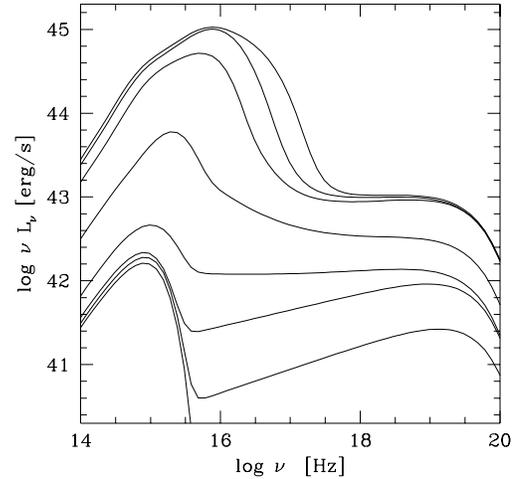, width=9.0cm, bbllx=0pt, bblly=210pt,
  bburx=624pt, bbury=680pt, clip=}
  \end{center}
  \caption{\em The AGN spectrum predicted by the accreting corona model
   for black hole mass $10^8 M\_{odot}$, 
   viscosity parameter $\alpha=0.33$ and values of $\LE$ ratio equal 
   0.00003, 0.00036, 0.000045, 0.001,
   0.01,0.1, 0.2 and 0.25, correspondingly}
  \label{fig:spec1}
\end{figure}

The first method was adopted for example by \cite*{bm86} and by \cite*{wy85}. 

In this paper we follow the second path used for example by \cite*{sm91}
but we use new theoretical model.
The model is a version of accretion disk with a hot corona based on
the assumption that accretion proceeds both through the disk and through 
the corona. Both gas phases are characterized by the viscosity parameter
$\alpha$ and the stratification of the flow is determined by atomic physics.
The model predicts both the optical/uv/soft X-ray emission (disk component)
as well as hard X-ray emission from the corona and it is parametrized by
the mass of the black hole, $M$, accretion rate, $\dot M$, and viscosity 
$\alpha$.

\section{ACCRETING CORONA MODEL}

\label{sec:model}

Since the publication of the paper of \cite*{hm91} accretion disk models with
hot optically thin thermal corona gained much attention. The model 
underwent significant developement in several directions, like
better description of the radiative transfer of X-rays 
and introdution of clumpiness. 

The version of the model proposed by \cite*{zcc95} and modified by 
\cite*{wcz96} differs qualitatively from the previous approach. At the expense
of assumption that the corona itself accretes and generates energy through 
viscosity we are able to predict the fraction of the
energy generated in the corona instead of adopting this quantity as a free
parameter of the model. The model therefore has considerable predictive 
power. In particular, it shows systematic change of the overall
spectra with the change of the accretion rate, i.e. the $\LE$ ratio.

The model is parametrized by the mass of the central black hole, $M$, 
accretion rate or $\LE$ ratio and the viscosity parameter $\alpha$, the
same in the disk and in the corona. At each
radius $r$ we
calculate the fraction of energy dissipated  in the main disk body 
and in the corona,
as described in Appendix D of \cite*{wcz96}. In this paper we neglect the
outflow from the corona predicted by the model as it requires separate carefull
study so the accretion rate is assumed to be constant throughout the disk.

We calculate the radiation spectrum emitted at each radius, $r$. 
Optically thick
disk emission is computed neglecting bound-free transitions and taking into
account the effects of electron scattering in a simplified way used by
\cite*{ce87}. The surface density of the disk is assumed to be equal 
0.1 of the mean disk density (if no corona exists at this radius) or 
equal the disk surface density determined by the corona pressure. 
This method gives results not much different from more advanced
computations of \cite*{drsr96}. Next, the local 
disk spectrum is Comptonized by the hot corona
of the optical depth, $\tau_{es}(r)$, and electron temeprature, $T_e(r)$, 
computed
at this radius from the model. We use for that purpose the simple 
approximation given by \cite*{cz91}. We assume that the disk albedo is equal
zero, i.e. we neglect the reflection component in the present study.

The final spectrum is computed by integrating 
over the disk surface. All the computations are done for non-rotating black
hole. We show a few examples of the spectra in Fig. \ref{fig:spec1}.

\begin{figure}
  \begin{center}
    \leavevmode
\epsfig{file=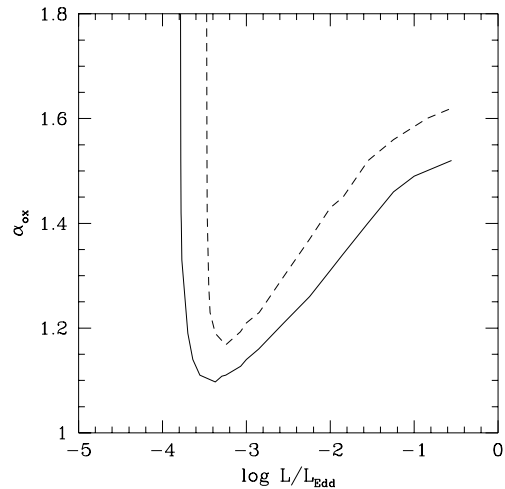, width=9.0cm, bbllx=0pt, bblly=210pt,
  bburx=624pt, bbury=680pt, clip=}
  \end{center}
  \caption{\em The dependence of the energy index $\alpha_{ox}$ from
  the $\LE$ ratio predicted by the accreting corona model for the mass of
  the black hole $10^8 M_{\odot}$ and two values of the viscosity parameter
   $\alpha$: 0.1 (continuous line) and 0.33 (dashed line)}
  \label{fig:alphaox_vs_lum}
\end{figure}

We see that an increase of accretion rate 
leads to the relative increase of the big blue 
bump bolometric luminosity with respect
to the bolometric luminosity in X-rays. This trend is a characteristic 
property of the model at any single radius above certain accretion rate 
treshold and reflects to some extent 
the behaviour of the model at $\sim 10 R_{Schw}$.
Below the treshold
value corona does not form and only disk emission is present. 
This treshold is an increasing function of the radius.  Therefore, the 
extension of the disk part covered by the corona increases with increasing
accretion rate and dominating part of X-ray emission comes from more and
more distant 
parts of the disk. Therefore, whilst disk emission at $\sim 10 R_{Schw}$
approximates well the UV/EUV part of the spectrum X-ray band (for larger
accretion rate) and optical band are dominated by emission from outer parts.

Another interesting prediction of the accreting corona model is the 
independence of the extension of the spectrum into high energies on any of the
model parameters. It is caused by the electron temperature in the corona 
being always $\sim 250 - 300 $keV at the radius where the corona is the 
strongest. This behaviour is easily seen from the analytic formulae given by
\cite*{wcz96}.

To express conveniently the predicted strength of the big blue bump with 
respect to the X-ray emission we
compute the energy index
$\alpha_{ox}$ between 2500~ \AA~ and 2 keV.
The result is ploted in 
Fig.\ref{fig:alphaox_vs_lum} for two values of the viscosity parameter 
$\alpha$. It depends on
the adopted value of the mass of the black hole (see Figs. \ref{fig:alphaox}
and \ref{fig:walfink}) due to the displacement of the big blue bump into
lower frequencies with an increase of the mass.

We can see that higher viscosity corresponds to values of $\LE$ ratio larger
by a factor $\sim 3$
for the same spectral slope $\alpha_{ox} = 1.4$.

\section{COMPARISON WITH THE DATA}

\subsection{Estimation of viscosity}

\begin{figure}
  \begin{center}
    \leavevmode
\epsfig{file=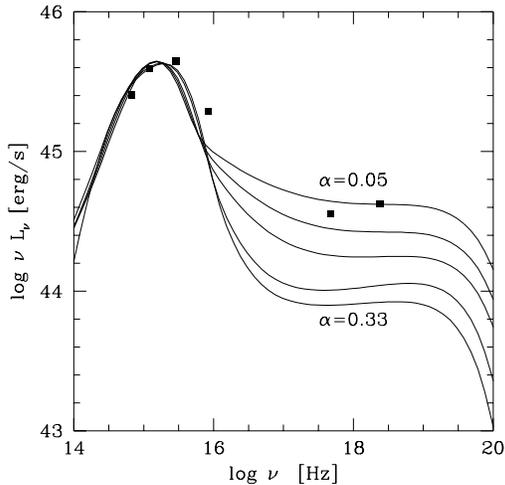, width=9.0cm, bbllx=0pt, bblly=210pt,
  bburx=624pt, bbury=680pt, clip=}
  \end{center}
  \caption{\em The AGN spectrum predicted by the accreting corona model
   for parameters adopted by Zheng et al. (1996) to represent a composite HST
  quasar spectrum (black hole mass $1.4 \times 10^9 M_{\odot}$, accretion rate
   $2.8 M_{\odot}/yr$). Curves were computed assuming viscosity parameter 
  $\alpha$ equal 0.5, 0.33, 0.2, 0.1 and 0.05. 
  Squares represent the composite spectrum of
  Zheng et al. (UV data) whilst optical/UV slope and UV/X-ray slope are from 
  Green (1996),  and hard X-ray slope is put to 0.9}
  \label{fig:visc}
\end{figure}

To fix the value of the viscosity parameter 
$\alpha$ we first model
the mean quasar spectrum shown in Fig. \ref{fig:visc}. The data do not
come from a single sample but they may nevertheless approximate well the 
actual distribution as the shape is not significantly different from other
samples (see  e.g. \cite{el+94}). 

We use the value of the mass of the black hole and accretion rate from  Fig.
10 of \cite*{zhe+96}. The Comptonization effect predicted by our model is
different from Comptonization required by \cite*{zhe+96} so the fit is never
perfect, independently from the viscosity but the overall optical/UV/X-ray
spectrum is reasonably well represented for viscosity $\alpha$ about 0.1 so
we adopt this value in any further computations.

Values of the mass of the black hole and accretion rate considerably different
(more than a factor 2) from the adopted above provide significantly worse
representation of the UV/EUV data, independently from the viscosity.

We have to stress, however, that the corona predicted by our model is optically
thin and therefore it cannot provide the mechanism to smear the Lyman edge 
if present in the disk emission. 

\subsection{Quasars}

We compute the predicted values of the spectral index $\alpha_{ox}$ for
a broad range of values of the masses of black holes and accretion rates.
The results are expressed in term of the rest frame luminosity 
$log(\nu L_{\nu})$ at 2500 \AA~ as this is the directly measurable quantity.

\begin{figure}
  \begin{center}
    \leavevmode
\epsfig{file=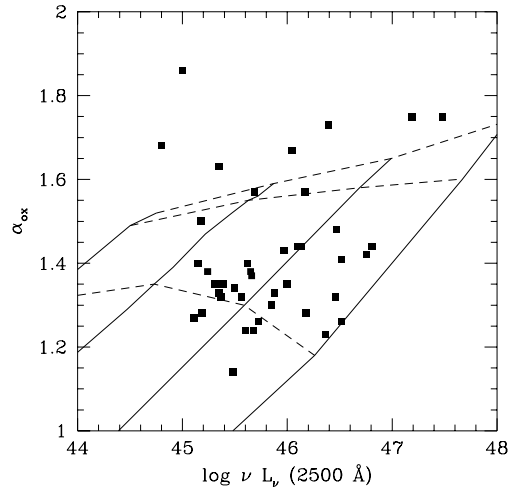, width=9.0cm, bbllx=0pt, bblly=210pt,
  bburx=624pt, bbury=680pt, clip=}
  \end{center}
  \caption{\em The dependence of the spectral index $\alpha_{ox}$ 
   (between 2500 \AA~ and 2 keV)
  on accretion rate
  calculated from the model assuming the viscosity parameter $\alpha=0.1$
  and the black hole mass equal $10^8, 10^9, 10^{10}$ and $10^{11} M_{\odot}$
  (continuous lines). Dashed curves mark the fixed $\LE$ ratio (0.01,0.1 and
  0.3).  Computations extend to 0.3 $\LE$ since beyond this limit 
   present spectra 
   are not reliable. Data points mark quasars from Green (1996)}
  \label{fig:alphaox}
\end{figure}

We compare these results against the data point taken from the sample of
\cite*{gre96}. Most data point group around the median values (45.95,1.38)
which correspond to central mass about $10^{10}$ and $\LE$ about 0.02. This
mass is higher than favored by \cite*{zhe+96} since this sample contains more
high luminosity objects. A few quasars have $\LE$ ratio higher than 0.3 but
we cannot extend our computations into that parameter space. Two extreme
object with the lowest luminosity may actually pose a problem to the model,
other objects can be reasonably well accomodated within it.

Most of the quasars, therefore, are not actually close to the Eddington limit
but occupy mostly the parameter space 0.008 - 0.1. It coincide well with
the behavior of the X-ray novae: the accretion is relatively stable and the
spectra are dominated by the big blue bump if the accretion rate is between
$\sim 0.01$ and a fraction of the Eddington luminosity.

\subsection{Seyfert galaxies}

\begin{figure}
  \begin{center}
    \leavevmode
\epsfig{file=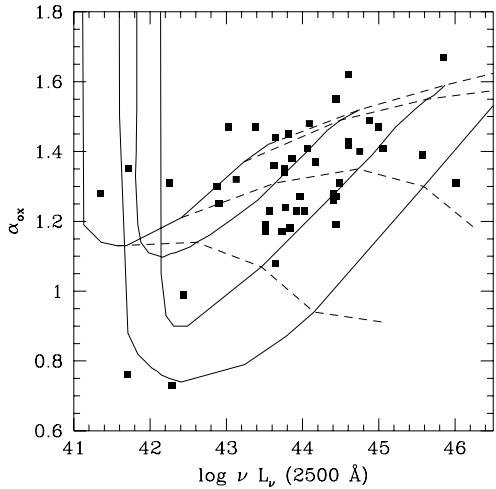, width=9.0cm, bbllx=0pt, bblly=210pt,
  bburx=624pt, bbury=680pt, clip=}
  \end{center}
  \caption{\em The dependence of the spectral index $\alpha_{ox}$ 
   (between 2500 \AA~ and 2 keV)
  on accretion rate
  calculated from the model assuming the viscosity parameter $\alpha=0.1$
  and the black hole mass equal $10^7, 10^8, 10^{9}$ and $10^{10} M_{\odot}$
  (continuous lines). Dashed curves show the fixed $\LE$ ratio (0.001. 
  0.01,0.1 and 0.3).
  Data points mark Seyfert galaxies from
  Walter \& Fink (1993).}
  \label{fig:walfink}
\end{figure}

Our sample of Seyfert galaxies is taken from \cite*{wf93}. They are shown
in Fig. \ref{fig:walfink}. Most data points
group around the median values (43.86,1.31) which corresponds to the central 
mass about $10^8 M_{\odot}$ and $\LE$ ratio about 0.01.

The objects are thus,
on average, much fainter but they actually cover a broad range of intrinsic
luminosities. The brighter objects join smoothly the distribution of quasars
mixing with them considerably without any clear systematic shift (see
Fig. \ref{fig:alphaox}). Also the $\LE$ ratio obtained for Seyferts is not
very much different from quasar values but it extends towards lower values
of $\LE$, mostly covering the range 0.001 - 0.1. It is therefore diffcult
to say that there is a systematic difference between the quasars and Seyferts
with respect to the $\LE$ ratios as the absence of weak blue bump quasars
might well be due to the selection effect. 

On the other hand, the extension of the observed distribution of Seyferts
towards $\LE$ ratios below 0.01 agrees well with the behaviour of X-ray novae
which are dominated by hard X-ray power low at low accretion rates and
show stronger variability.

\subsection{Narrow Line Seyfert galaxies}

Steep Spectrum Seyfert galaxies, or Narrow Line Syfert 1 galaxies 
consitute some 10 - 15 \% of the Seyfert galaxy populations (e.g. 
\cite{pb96}). They are so strongly dominated by the big blue bump  component
that it is difficult to estimate any contribution from the standard hard
X-ray power law with index $\sim 0.9 - 1.0$. Looking at 
Fig.\ref{fig:alphaox_vs_lum} we see that two types of objects are dominated
by big blue bump. 

The first class are objects close to the Eddington limit. 
Unfortunately our computations of the spectra do not extend beyond 
$\LE \sim 0.3$ so we cannot predict the spectral shapes quantitatively. 
However, we expect that independently from the luminosity the spectra should
show the hard tail and this hard X-ray emission should not vary strongly
since it comes from large radii and only some local instabilities within
the corona (if there are any) may influence the level of emission. However, the
situation may be more complex very close to the Eddington limit since the
behavior of the gas may ceaze to be stationary (\cite{wcz96}).

The second class are objects with very small $\LE$ ratio. 
In these object corona forms only in the innermost part of the disk or it 
is entirely absent, according to the present model. Small variations in
accretion rate can produce enormous effect if we are close to the transitory
accretion rate value.

The two classes of object differ strongly with respect to the peak frequency
of the big blue bump for given luminosity. Since NL Sy1 extends well into
soft X-rays the first option, i.e. large $\LE$ ratio, is clearly favored.

\section{CONCLUSIONS}

The model of accreting corona well reproduces the overall optical/UV/X-ray
spectra in AGN. It shows some curvature in soft X-ray bend consistent with
spectra being steeper below 2 keV. The high energy extension of the spectra 
)$\sim 250 - 300 $ keV is independent from model parameters.

Comparison of the data with the optical/UV/X-ray spectra predicted by the
this model show that quasars radiate usually at $\sim 0.01-0.1$
of the Eddington luminosity, Seyfert galaxies at $\sim 0.001 - 0.3$ and Narrow
Line Seyfert galaxies are close to the Eddington limit.

\section*{ACKNOWLEDGMENTS}

We thank Pawel Czerny for typing in the data points.
This work was supported in part by grants No. 2P03D00410 and 2.P03C.005.11p/01
of the Polish State 
Committee for Scientific Research.

\clearpage


\end{document}